\documentclass[10pt,letterpaper]{article}
\pdfoutput=1

\usepackage{opex3}

\usepackage[cmex10]{amsmath}
\usepackage{amssymb}
\usepackage{subfigure}
\usepackage{cite} 

\definecolor{darkviolet}{rgb}{0.58, 0.0, 0.83}
\def\a{s}
\def\b{s}
\newcommand{\add}[1]{\if\a\b{{\color{red} #1}}\else{#1}\fi} 
\newcommand{\del}[1]{{\if\a\b{{\color{blue}[[#1]]}}\else{}\fi}}
\newcommand{\com}[1]{\if\a\b{\textbf{\color{darkviolet} #1}}\else{}\fi}
 
\begin{document}

\title{SIMPEL: Circuit model for photonic spike processing laser neurons}

\author{Bhavin J. Shastri, Mitchell A. Nahmias, Alexander N. Tait, Ben Wu, and Paul R. Prucnal}

\address{Princeton University, Princeton NJ, 08544, USA}

\email{shastri@ieee.org} 



\begin{abstract}
We propose an equivalent circuit model for photonic spike processing laser neurons with an embedded saturable absorber---a simulation model for photonic excitable lasers (SIMPEL). We show that by mapping the laser neuron rate equations into a circuit model, SPICE analysis can be used as an efficient and accurate engine for numerical calculations, capable of generalization to a variety of different laser neuron types found in literature. The development of this model parallels the Hodgkin--Huxley model of neuron biophysics, a circuit framework which brought efficiency, modularity, and generalizability to the study of neural dynamics. We employ the model to study various signal-processing effects such as excitability with excitatory and inhibitory pulses, binary all-or-nothing response, and bistable dynamics.
\end{abstract}



\bibliographystyle{./osajnl}
\bibliography{./References}

\section{Introduction}\label{sect:introduction}
Photonics has recently witnessed\cite{selmi_2014,brunner_2013,woods_2012,martinenghi_2012,hurtado_2012,appeltant_2011,nahmias_jstqe_2013,coomans_2011,kravtsov_2011,vandoorne_2008,Romeira:13,VanVaerenbergh:13} a deeply committed exploration of neuro-inspired unconventional computing paradigms that promise to outperform conventional technology in certain problem domains. This field of neuromorphic engineering\cite{jaeger_2004,hasler_2013,mahowald1991silicon} aims to build machines that better interact with natural environments by applying the circuit and system principles of neuronal computation, including robust analog signaling, physics-based dynamics, distributed complexity, and learning. Cognitive computing platforms\cite{Merolla08082014,modha_2011} inspired by the architecture of the brain promise potent advantages in efficiency, fault tolerance and adaptability over von Neumann architectures for tasks involving pattern analysis, decision making, optimization, learning, and real-time control of many-sensor, many-actuator systems. These neural-inspired systems are typified by a set of computational principles, including hybrid analog-digital signal representations\cite{sarpeshkar_1998}, co-location of memory and processing\cite{szatmary_2010}, unsupervised statistical learning\cite{abbott00}, and distributed representations of information\cite{Maass2002a}.

A sparse coding scheme called \emph{spiking}\cite{izhikevich_2003,thorpe2001spike} has been recognized as a cortical encoding strategy\cite{ostojic_2014,diesmann_1999} with firm code-theoretic justifications\cite{kumar_2010,borst_1999} and promises extreme improvements to computational power efficiency\cite{hasler_2013}. Since spikes are discrete events that occur at analog times, this encoding scheme represents a hybrid between traditional analog and digital approaches, capable of both expressiveness and robustness to noise \cite{tait2014photonic}. This distributed, asynchronous model processes information using both space and time\cite{thorpe2001spike,Maass2002a}, is naturally robust, and is amenable to algorithms for unsupervised adaptation\cite{abbott00,szatmary_2010}. The marriage of photonics with spike processing is fundamentally enabled by the strong analogies\cite{nahmias_jstqe_2013} of the underlying physics between the dynamics of biological neurons and lasers; they both can be understood within the framework of dynamical systems theory, and can display the crucial property of excitability---systems that can be excited from their in a stable steady rest state to emit a spike by a super-threshold followed by a refractory period. The rate equations of photonics devices, however operate approximately eight orders of magnitude faster than biological time scales. In addition to the high switching speeds and high communication bandwidth, the low cross-talk achievable in photonics are very well suited for an ultrafast spike-based information scheme with high interconnection densities. Furthermore, the high wall-plug efficiencies of photonic devices may allow such implementations to match or eclipse equivalent electronic systems in low energy usage. Consequently, a network of photonic neurons---photonic spike processors---could access a \emph{picosecond} and \emph{low-power} computationally rich domain that is inaccessible by other technologies. This novel processing domain---\emph{ultrafast cognitive computing}---represents a broad domain of applications where quick, temporally precise and robust systems are necessary, including: adaptive control, learning, perception, motion control, sensory processing, autonomous robotics, and cognitive processing of the radio frequency (RF) spectrum.

Despite their potential in this regard, all the excitable laser systems studied in the context of spike processing, either with the tools of bifurcation theory \cite{Alexander:13,coomans_2011,shastri_nusod_2013} or experimentally \cite{hurtado_2012,shastri_pdp_ipc_2013}, have been limited to a couple of devices. This has been largely due to the lack of a dynamical simulation platform for photonic neurons. Due to the complexity associated with these systems, it is impossible to apply simplifying assumptions for purely analytic approaches which would risk not capturing all the rich dynamics associated with such systems. A platform for simulating photonics neurons must simultaneously capture their underlying physics (photon--carrier dynamics) and be amenable for studying large scale on-chip optical network architecture to support massively parallel communication between high-performance spiking laser neurons \cite{tait_jlt_2014,tait_OI_2014,nahmias_OI_2014} (also highlighted in \cite{shastri_newsletter_2014}). Furthermore, such a simulation platform would be crucial to gain critical insight into the behavior of a photonic spike processor under a variety of conditions.

The photon--carrier dynamics in a excitable laser are strongly coupled, making the simulation of transient spiking phenomena in laser neurons much more involved than steady-state simulation of typical CW lasers. Multidimensional device-level programs (for example, based on finite element method (FEM)) are too computationally intensive to accurately account for these complex photon--carrier interactions, and they are also impossible to utilize for simulating a scalable integrated optoelectronic system with hundreds or thousands of such devices. Numerical methods such as the Runge--Kutta methods can be employed for calculating solutions of the rate equations. However, they have a number of limitations for modeling practical situations; for example, they cannot easily incorporate parasitic networks for studying device interactions and determining the frequency response. Other simulation techniques take advantage of the sparse nature of spiking signals and adopt event-based models for extremely large-scale networks. While computationally efficient, event-based models must overlook many subtleties in the dynamics of individual units, which are fundamentally driven by differential equation models.

\subsection{Our contribution}
Here, we propose \emph{SIMPEL}---SImulation Model for Photonic Excitable Lasers---which bridges the gap between the underlying physics and relevant dynamics of excitable lasers by transforming its rate equations to an equivalent circuit representation. We show that this circuit model leads to a highly efficient simulation framework that expedites the computation of the rate equations, and also facilitates computer-aided design (CAD) and analysis of an integrated optoelectronic system. The circuit modeling approach captures the important dynamical behaviors of a photonic neuron in a way that is extendable to more refined studies of parasitics and networks while also being agnostic to specifics of device implementation. In our approach, the carrier density and the photon density are converted to current and voltage representations, respectively. The input current (carrier) and output voltage (light) form a two-port equivalent model. This two-port circuit can be coupled with other such circuits for an efficient and scalable simulation platform to study for example an architecture of an interconnected network of laser neurons at the system level for wide range of applications in high-performance computing, adaptive control, and learning. Our model is compatible with SPICE, the de facto industrial standard for computer-aided circuit analysis, which is hyper optimized for circuit analysis.

By recasting excitable laser dynamics in a common language of circuit analysis, SIMPEL manifests several advantageous features that will aid in the design and characterization of laser neuron devices for signal processing. This approach parallels the work of Hodgkin and Huxley\cite{hodgkin_huxley}, who developed an equivalent circuit model for biological neuron dynamics in order to express neuron behavior in a common language of engineering that can abstract away underlying biochemical mechanisms. A circuit analysis framework is a modular abstraction, which is important for incorporating second-order effects, such as parasitics, and for studying small networks of interconnected laser devices. Compatibility with the SPICE engine means that the simulation is highly optimized. The resulting increase in speed over physical and Runge--Kutta methods is particularly relevant for Monte--Carlo studies of noise in which stochastic effects are studied with many calls of the same simulation. Finally, SIMPEL's phenomenological (a.k.a. behavioral) nature makes it applicable to different physical implementations of an excitable laser with embedded saturable absorber (SA) without resorting to full-scale device simulation. Although the model and parameters must be derived from the underlying optoelectronic mechanisms, SIMPEL can more readily grant insight about the relationship between these parameters and behavioral dynamics relevant for signal processing, such as energy usage and noise. One example of an application is parameter fitting data measured from a laser neuron circuit. In this case, the effects of noise and parasitics must be accounted for, and the internal physical mechanism is not observable. SIMPEL could allow a user to learn the most about circuit operation by improving control over the fit degrees-of-freedom and, more generally, over the synthesis of \textit{a priori} knowledge with observed behavior.

\begin{figure*}[t]
	\centering\includegraphics[width=5in]{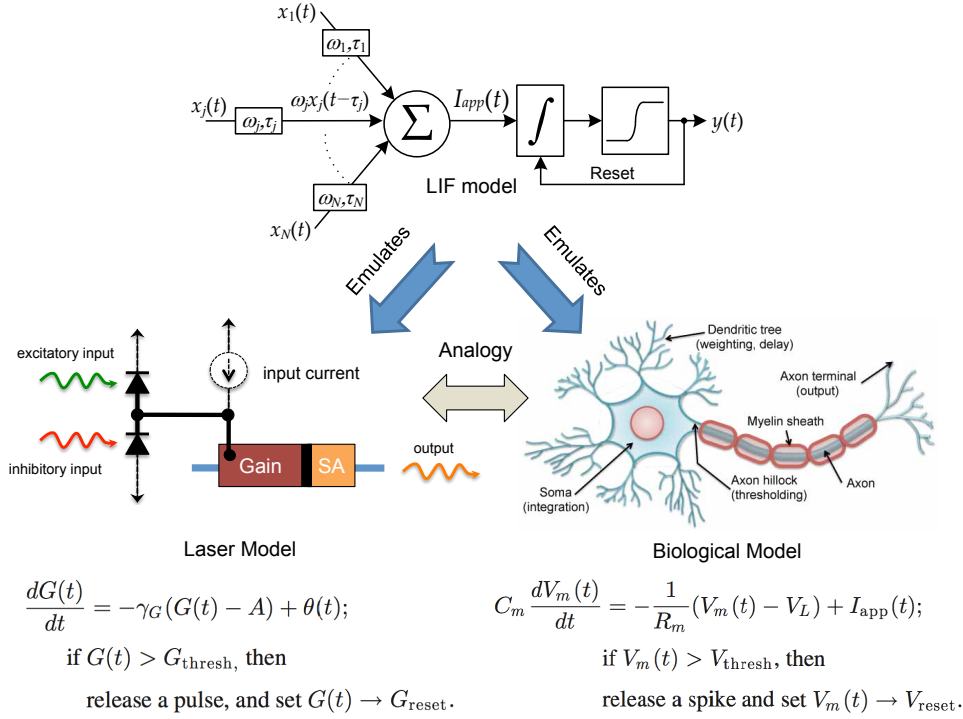}
	\caption{Schematic of a biological neuron and a two-section excitable laser that share key dynamical properties. In the LIF neuron model, weighted and delayed input signals are spatially summed at the dendritic tree into an input current, which travel to the soma and perturb the internal state variable, the voltage. The soma performs integration and then applies a threshold to make a spike or no-spike decision. After a spike is released, the voltage is reset. The resulting spike is sent to other neurons in the network. The excitable laser is composed of a gain section, SA, and mirrors for cavity feedback. The inputs selectively perturb the gain optically or electrically. The gain medium acts as a temporal integrator while the SA acts as a threshold detector; it extracts most of the stored energy from the gain medium into the optical mode. These dynamics emulate \emph{excitability}, one of the most critical properties of a spiking neuron.}
	\label{fig:analogy}
\end{figure*}

\section{Brief review of spiking neuron model and excitable laser model analogy}\label{sect:background}
\subsection{Spiking neuron model}
Fig.\,\ref{fig:analogy} depicts the basic biological structure of the leaky integrate-and-fire (LIF) neuron. The LIF model is one of the most ubiquitous models in computational neuroscience and is the simplest known model for spike processing\cite{izhikevich2004model}. The dendrite tree collects $N$ inputs which represent induced currents in input synapses $x_j$ that are continuous time series consisting either of spikes or continuous analog values. Each input is independently weighted by $\omega_j$, which can be positive or negative, and delayed by $\tau_j$ resulting in a time series that is spatially summed. This aggregate input induces an electrical current, $I_\mathrm{app}=V_m(t)\sum_{j=1}^{N}\omega_jx_j(t-\tau_j)$ between adjacent neurons, where the membrane potential $V_m(t)$, the voltage difference across their membrane, acts as the primary internal (activation) state variable. The weights and delays determine the dynamics of network, providing a way of programming a neuromorphic system. The soma acts as a first-order low-pass filter or a leaky integrator, with the integration time constant $\tau_m=R_mC_m$ that determines the exponential decay rate of the impulse response function and where $R_m$ and $C_m$ model the resistance and capacitance associated with the membrane, respectively. The leakage current through $R_m$ drives the membrane voltage $V_m(t)$ to 0, but an active membrane pumping current counteracts it and maintains a resting membrane voltage at a value of $V_m(t)=V_L$; that is $V_m(t)=V_L e^{-\frac{t-t_0}{\tau_m}}+\frac{1}{C_m}\int_{0}^{t-t_0}I_\mathrm{app}(t-\tau)e^{-\frac{\tau}{\tau_m}}$, where $t_0$ is the last time the neuron spiked. Finally, the axon carries an action potential, or spike, to other neurons in the network when the integrated signal exceeds a threshold; that is, if $V_m(t)\geq V_\mathrm{thresh}$, then the neuron outputs a spike and $V_m(t)$ is set to $V_\mathrm{reset}$. This is followed by a relative \emph{refractory period}, during which $V_m(t)$ recovers from $V_\mathrm{reset}$ to the resting potential $V_L$ in which is difficult to induce the firing of a spike. Consequently, the output of the neuron consists of a series of spikes that occur at continuously valued times. There are three influences on $V_m(t)$: passive leakage of current, an active pumping current, and external inputs generating time-varying membrane conductance changes. Including a set of digital conditions, we arrive at a typical LIF model for an individual neuron:
\begin{subequations}
\begin{align}
	\underbrace{\frac{dV_m(t)}{dt}}_\textrm{Activation}&=\underbrace{\frac{V_L}{\tau_m}}_\textrm{Active pumping}-\underbrace{\frac{V_m(t)}{\tau_m}}_\textrm{Leakage}+\underbrace{\frac{1}{C_m}I_{app}(t)}_\textrm{External input};\\
	&\text{if $V_m(t)>V_{\mathrm{thresh}}$ then}\\\nonumber
	&\text{release a pulse and set $V_m(t)\rightarrow V_{\mathrm{reset}}$.}
\end{align}
\label{eq:LIF}
\end{subequations}

\subsection{Excitable laser model}
Next, we briefly summarize the recently discovered mathematical analogy between the LIF neuron model and an excitable laser composed of a gain section with an embedded SA \cite{nahmias_jstqe_2013,nahmias_ipc_2013} as illustrated in Fig.\,\ref{fig:analogy}. The gain medium acts as a temporal integrator with a time constant that is equal to the carrier recombination lifetime. The SA extracts most of the stored energy from the gain medium into the optical mode and performs the function of a threshold detector. This gain-absorber interplay emulates one of the most critical dynamical properties of a spiking neuron---excitability.

The Yamada model \cite{dubbeldam_1999}, describes the behavior of lasers with independent gain and SA sections with an approximately constant intensity profile across the cavity. We assume that the dynamics operate such that the gain is a slow variable, while the intensity and loss are both fast. This three-dimensional dynamical system can be described with the following equations: (1) $\dot{G}(t)=\gamma_G\left[A-G(t)-G(t)I(t)\right]$; (2) $\dot{Q}(t)=\gamma_Q\left[B-Q(t)-aQ(t)I(t)\right]$; and (3) $\dot{I}(t)=\gamma_I\left[G(t)-Q(t)-1\right]I(t)+\epsilon f(G)$ where $G(t)$ models the gain, $Q(t)$ is the absorption, $I(t)$ is the laser intensity, $A$ is the bias current of the gain, $B$ is the level of absorption, a describes the differential absorption relative to the differential gain, $\gamma_G$ is the relaxation rate of the gain, $\gamma_Q$ is the relaxation rate of the absorber, $\gamma_I$ is the reverse photon lifetime, and $\epsilon f(G)$ represents the small contributions to the intensity made by spontaneous emission, (noise term) where $\epsilon$ is very small.

It has been shown in \cite{nahmias_jstqe_2013} that, in certain parameter regimes, the behavior of the system closely approximates the spiking LIF model. Assuming that the inputs to the system cause perturbations to the gain $G(t)$ only, and that the fast dynamics are nearly instantaneous, we can compress the behavior of this system into the following set of equations and conditions:
\begin{subequations}
\begin{align}
	\underbrace{\frac{dG(t)}{dt}}_\textrm{Activation}&=\underbrace{\gamma_GA}_\textrm{Active pumping}-\underbrace{\gamma_GG(t)}_\textrm{Leakage}+\underbrace{\theta(t)}_\textrm{External input}\\
	&\text{if $G(t)>G_{\mathrm{thresh}}$ then}\\\nonumber
	&\text{release a pulse, and set $G(t)\rightarrow G_{\mathrm{reset}}$.}
\end{align}
\end{subequations}
where $\theta(t) $ represent input perturbations. The conditional statements account for the fast dynamics of the system that occur on times scales of order $1/\gamma_{I}$, and other various assumptions---including the fast $Q(t)$ variable and operation close to threshold---assure that $G_{\mathrm{thresh}}$, $G_{\mathrm{reset}}$ and the pulse amplitude remain constant.

Comparing this to the LIF model, or equation \eqref{eq:LIF}, the analogy between the equations becomes clear. Setting the variables $\gamma_G=1/R_m C_m$, $A=V_L$, $\theta(t)= I_{app}(t)/R_m C_m$, and $G(t)=V_m(t)$ shows their algebraic equivalence. Thus, the gain of the laser $G(t)$ can be thought of as a virtual \emph{membrane voltage}, the input current $A$ as a virtual \emph{leakage voltage}, etc. There is a key difference, however---both dynamical systems operate on vastly different time scales. Whereas biological neurons have time constants $\tau_m=C_m R_m$ on order of milliseconds, carrier lifetimes of laser gain sections are typically in the nanosecond range and can go down to picosecond.

\section{Laser neuron rate equation model}\label{sect:laser_model}
\subsection{Rate equations}
We start with the coupled rate equations for the photon and carrier densities for a two-section excitable laser with gain and SA regions, assuming only one longitudinal and one transverse optical mode is lasing \cite{nugent_1995}:
\begin{eqnarray}
	\frac{dn_a}{dt}&=&\underbrace{\frac{\eta_{i,a} i_a}{qV_a}}_\text{current injection}- \underbrace{\frac{n_a}{\tau_a}}_\text{carrier recomb.}-\underbrace{\Gamma_ag(n_a)\frac{N_{ph}}{V_a}}_\text{stimulated-emission}\\
	\frac{dn_s}{dt}&=&\underbrace{\frac{\eta_{i,s} i_s}{qV_s}}_\text{current injection}-\underbrace{\frac{n_s}{\tau_s}}_\text{carrier recomb.}-\underbrace{\Gamma_sg(n_s)\frac{N_{ph}}{V_s}}_\text{stimulated-emission}\\
	\frac{dN_\mathrm{ph}}{dt}&=&-\underbrace{\frac{N_{ph}}{\tau_{ph}}}_\text{photon decay}+\underbrace{\Gamma_ag(n_a)N_{ph}+\Gamma_sg(n_s)N_{ph}}_\text{stimulated-emission} + \underbrace{V_a\beta B_r n_a^2}_\text{recombination}
\end{eqnarray}
Equations (3) and (4) relate the rate of change in the gain and SA regions' carrier concentration $n_\chi$ to the injection current $i_\chi$, the carrier recombination rate, and the stimulated-emission rate. Note that the gain and SA cavities are represented by the subscript $\chi=a$ or $s$, respectively. The gain current term $i_a = I_a + i_{ea}$ accounts for the pump current $I_a$ and the electrical modulation $i_{ea}$ of the gain, whereas the SA current term $i_s = I_s$ allows only for an adjustable threshold. Equation (5) relates the rate of change in photon number $N_\mathrm{ph}$ that is common to the gain and SA regions, to photon decay, the rate of stimulated-emission, and the rate of recombination into the lasing mode. In addition, $\eta_{i,\chi}$ is the current-injection efficiency, $V_\chi$ is the cavity volume, $q$ is the electron charge, $\tau_\chi$ is the rate of carrier recombination, $\Gamma_\chi$ is the optical confinement factor, $\beta$ is the spontaneous-emission coupling factor, $B_r$ is the bimolecular recombination term, and $\tau_\mathrm{ph}$ is the photon lifetime given as $\tau_\mathrm{ph}^{-1}=(c/n_r)[\alpha+\ln(1/R_1R_2)/2L]$, where $R_1$ and $R_2$ are the cavity mirror reflectivities, $L$ is the cavity length, and $\alpha$ is the internal loss of the cavity. Note that, for the active and passive regions to stay as gain and absorber regions, $I_a>qV_an_a/\tau_a$ and $I_s<qV_sn_s/\tau_s$, respectively. The stimulated-emission rate includes a carrier-dependent gain term $g(n_\chi)$ which can take on either a linear \cite{javro_1995} or logarithmic form \cite{detemple_1993}. We chose the former for simplicity; that is, $g(n_\chi)=g_\chi(n_\chi-n_{0,\chi})$, where $g_\chi$ is the gain and SA region differential gain and loss coefficient, respectively, and $n_{0,\chi}$ is the optical transparency carrier density. Furthermore, these rate equations can be further generalized by adding a gain-saturation term $\phi_\chi^{-1}(N_\mathrm{ph})=1/(1+\epsilon_\chi\Gamma_\chi N_\mathrm{ph})$ \cite{channin_1979}, where $\epsilon_\chi$ is the phenomenolgical gain-compression factor. Finally, the laser output power $P_\mathrm{out}$ is related to the photon number inside the cavity via:
\begin{equation}
	\frac{N_\mathrm{ph}}{P_\mathrm{out}}=\frac{\lambda\tau_\mathrm{ph}}{\eta_c\Gamma_a h c}\triangleq\vartheta
\end{equation}
where $\lambda$ is the lasing wavelength, $\eta_c$ is the output power coupling coefficient, $h$ is Planck's constant, and $c$ is the speed of light in a vacuum.

\subsection{Equivalent circuit}
For a given set of injection currents $\{I_a,I_s\}$ in the gain and SA regions, operating point analysis of the excitable laser described by the rate equations (3)--(5) and output power (6) leads to four solutions. In addition to the correct nonnegative solution regime, in which the solutions for the photon density $N_\mathrm{ph}$, and carrier densities $n_a$ and $n_s$, are all nonnegative when $I_a\geq 0$ and $I_s\geq 0$, there are also negative-power and a high-power regimes. It is therefore necessary to transform the carrier population density in the respective cavities $n_\chi$ and the laser output power $P_\mathrm{out}$ via the following pair of transformations, respectively\cite{javro_1995}:
\begin{eqnarray}
	n_a&=&n_{\mathrm{eq},a}\exp\left(\frac{qv_a}{nkT}\right)\\
	n_s&=&n_{\mathrm{eq},s}\exp\left(\frac{qv_s}{nkT}\right)\\
	P_\mathrm{out}&=&(v_m+\delta)^2
\end{eqnarray}
where $n_{\mathrm{eq},\chi}$ is the equilibrium carrier density, $v_\chi$ is the voltage across the gain and SA region of the laser, $n$ is a diode ideality factor (typically set to two for III-V devices \cite{park_bowers_pd_2007,tucker_1983}), $v_m$ is a new variable for parameterizing $P_\mathrm{out}$, $\delta$ is a small arbitrary constant set to $10^{-60}$, $k$ is Boltzmann's constant, and $T$ is the temperature of the excitable laser. It has been shown in \cite{mena_1997} that these transformations eliminate the nonphysical solutions (negative-power and a high-power regime) and improve the convergence properties of the model during simulation.

\begin{figure*}[t]
	\centering
	\includegraphics[width=5.5in]{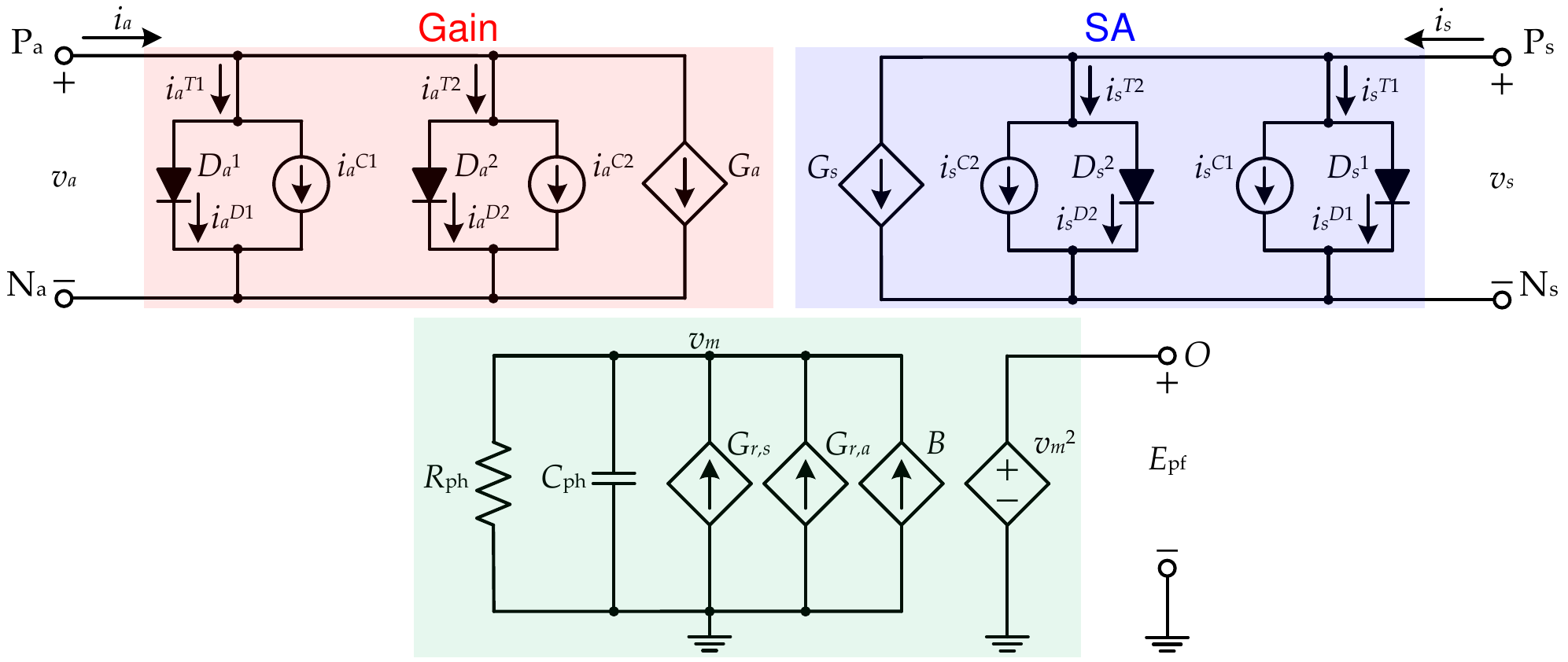}
	\caption{Circuit-level implementation to model the laser neuron.}
	\label{fig:circuit_model}
\end{figure*}

We model the carrier's dynamics $dn_\chi/dt$, by substituting the set of variable transformations (7)--(9), and the output power (6), into the rate equations (3) and (4). After applying the appropriate manipulations, we obtain:
\begin{equation}
	\frac{q n_{\mathrm{eq},\chi}}{nkT}\exp\left(\frac{qv_\chi}{nkT}\right)\frac{dv_\chi}{dt}=\frac{\eta_{i,\chi} i_\chi}{qV_\chi}-\frac{n_{\mathrm{eq},\chi}}{\tau_\chi}\left[\exp\left(\frac{qv_\chi}{nkT}\right)-1\right]-\frac{n_{\mathrm{eq},\chi}}{\tau_\chi}-\frac{\Gamma_\chi g(n_\chi)}{V_\chi}\vartheta(v_m+\delta)^2.
\end{equation}
With some additional rearrangement (10) can be written in terms of the respective cavity currents as
\begin{equation}
	i_\chi=i_\chi^{T1}+i_\chi^{T2}+G_\chi
\end{equation}
where
\begin{eqnarray}
	i_\chi^{T1}&=&i_\chi^{D1}+i_\chi^{C1}\\
	i_\chi^{T2}&=&i_\chi^{D2}+i_\chi^{C2}\\
	i_\chi^{D1}&=&\frac{q n_{\mathrm{eq},\chi} V_\chi}{2\eta_{i,\chi}\tau_\chi}\left[\exp\left(\frac{q v_\chi}{nkT}\right)-1\right]\\
	i_\chi^{D2}&=&\frac{q n_{\mathrm{eq},\chi} V_\chi}{2\eta_{i,\chi}\tau_\chi}\left[\exp\left(\frac{q v_\chi}{nkT}\right)-1+\frac{2q\tau_\chi}{nkT}\exp\left(\frac{qv_\chi}{nkT}\right)\frac{dv_\chi}{dt}\right]\\
	i_\chi^{C1}&=&\frac{q n_{\mathrm{eq},\chi} V_\chi}{2\eta_{i,\chi}\tau_\chi}\\
	i_\chi^{C2}&=&i_\chi^{C1}\\
	G_\chi&=&\frac{\vartheta q \Gamma_\chi}{\eta_{i,\chi}}g\left(\Theta_\chi i_\chi^{T1}\right)
\end{eqnarray}
with
\begin{equation}
	n_\chi = \Theta_\chi i_\chi^{T_1} \quad \text{and} \quad \Theta_\chi=\frac{2\eta_{i,\chi}\tau_\chi}{qV_\chi}.
\end{equation}
These equations can be mapped directly into an equivalent laser neuron model as shown in Fig.\,\ref{fig:circuit_model}, where $\{\mathsf{P_a},\mathsf{N_a}\}$ and $\{\mathsf{P_s},\mathsf{N_s}\}$ are the electrical ($+$ve and $-$ve) terminals of the gain and SA regions, respectively. Diodes $\{D_a^1,D_a^2\}$ and $\{D_s^1,D_s^2\}$, and current sources $\{i_a^{C1},i_a^{C2}\}$ and $\{i_s^{C1},i_s^{C2}\}$, model the linear recombination and charge storage in both the gain and SA regions, respectively. The nonlinear dependent current sources $\{G_a,G_s\}$, model the effect of stimulated emission on the carrier densities in both the gain and SA regions.

Similarly, to model the photon dynamics $dN_\mathrm{ph}/dt$, we substitute the transformations (7)--(9), and the output power (6), into the rate equation (5). After applying the appropriate manipulations, we obtain
\begin{equation}
	2(v_m+\delta)\frac{dv_m}{dt}=-\frac{(v_m+\delta)^2}{\tau_\mathrm{ph}}+\Bigl\{\Gamma_a g(n_a)+\Gamma_s g(n_s)\Bigr\}(v_m+\delta)^2+\frac{V_a\beta B_r n_a^2}{\vartheta}
\end{equation}
With some additional rearrangements and the definition of suitable circuit elements, (18) can be written as
\begin{equation}
	C_\mathrm{ph}\frac{dv_m}{dt}+\frac{v_m}{R_\mathrm{ph}}=G_{r,a}+G_{r,s}+B
\end{equation}
where
\begin{eqnarray}
	G_{r,a}&=&\tau_\mathrm{ph}\Gamma_a g\left(\Theta_a I_a^{T1})(v_m+\delta\right)-\delta\\
	G_{r,s}&=&\tau_\mathrm{ph}\Gamma_s g\left(\Theta_s I_s^{T1})(v_m+\delta\right)\\
	B&=&\frac{\eta_c \Gamma_a hc V_a \beta B_r}{\lambda(v_m+\delta)}\left(\Theta_a i_a^{T_1}\right)^2
\end{eqnarray}
and
\begin{equation}
	C_\mathrm{ph}=2\tau_\mathrm{ph} \quad \text{and} \quad R_\mathrm{ph}=1\Omega.
\end{equation}
Finally, $E_\mathrm{pf}$ transforms the node voltage $v_m$, into the output power $P_\mathrm{out}$, by
\begin{equation}
	P_\mathrm{out}=E_\mathrm{pf}=(v_m+\delta)^2.
\end{equation}
These equations can also be mapped directly into an equivalent laser neuron model as shown in Fig.\,\ref{fig:circuit_model}. $C_\mathrm{ph}$ and $R_\mathrm{ph}$ help model the time-variation of the photon density under the effect of spontaneous and stimulated emission, which are accounted for by the nonlinear dependent current sources $\{G_{r,a},G_{r,s}\}$ and $B$, respectively. Finally, the nonlinear voltage source $E_\mathrm{pf}$, produces the laser neuron optical output power in the form of a node voltage at $o$.

\subsection{Laser neuron device structures}
\begin{figure*}[t]
	\centering
	\subfigure[]{\includegraphics[width=2.15in]{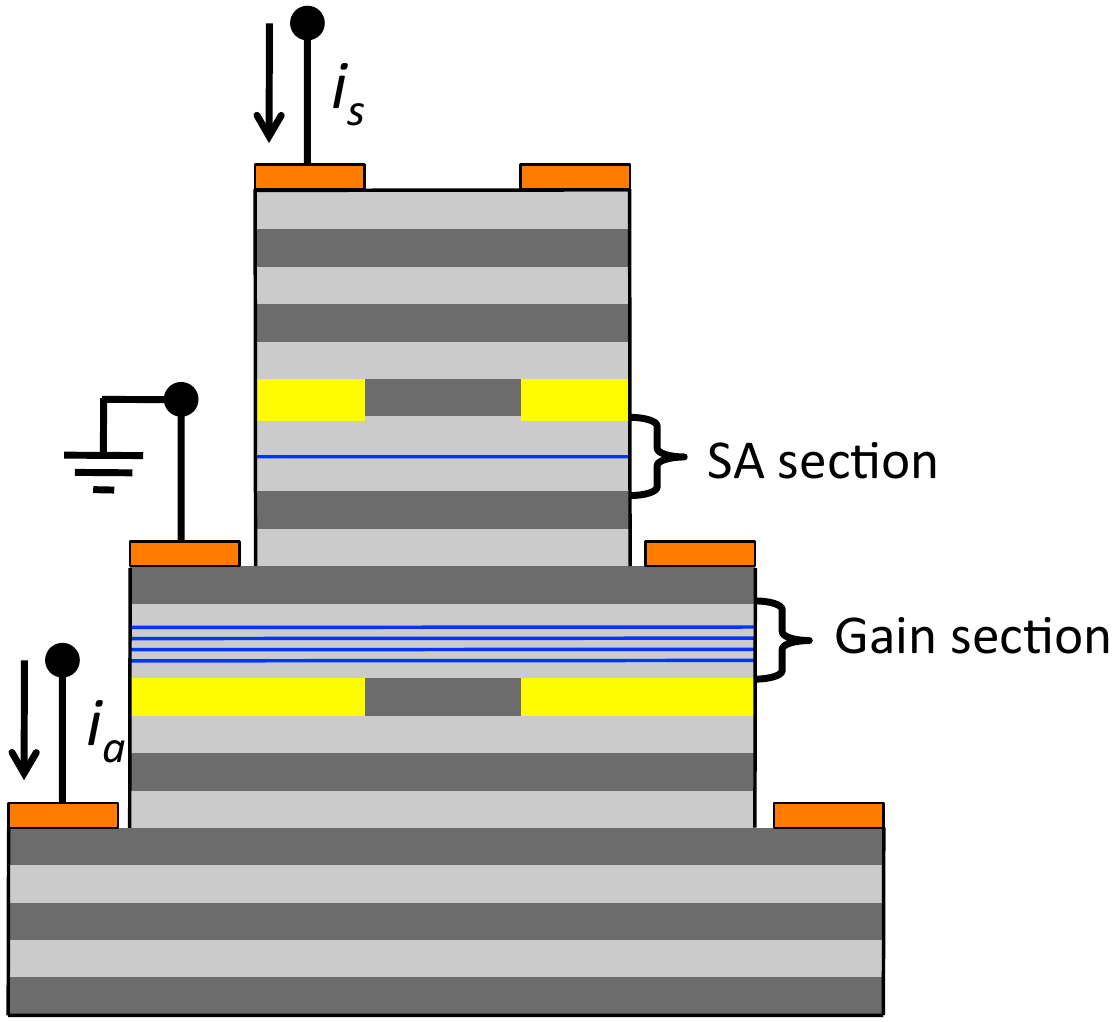}%
		\label{fig:vcsel_cross_section}}
	\subfigure[]{\includegraphics[width=2.85in]{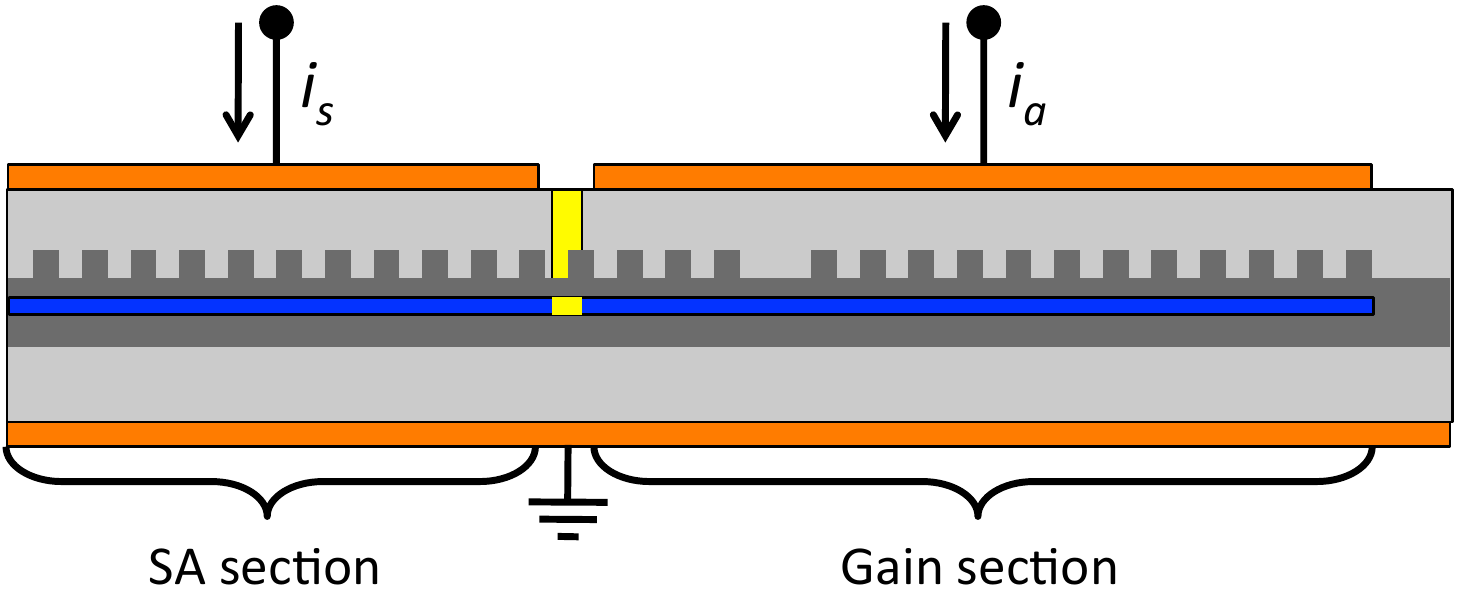}%
		\label{fig:dfb_cross_section}}
	\caption{(a) Cross section of a VCSEL-SA excitable laser. The single cavity mode interacts with two independent sets of quantum wells (blue). Metal contacts (orange) inject current into gain and SA sections from below and above, respectively. Insulating regions confine current to the mode center (yellow). (b) Cross section of a DFB-SA excitable laser. A refractive index grating etched in the waveguide core (dark grey) creates a high-finesse, single mode cavity whose mode interacts with both gain and SA MQW active regions (blue). Metal contacts (orange) inject current into gain and SA sections, which are electrically isolated by a H\textsuperscript{+} implanted layer (yellow).}
	\label{fig:vcsel_dfb_cross_sections}
\end{figure*}

As a demonstration of the utility of our circuit approach, we simulate our model using realistic parameters for the recently proposed vertical-cavity surface-emitting laser (VCSEL) photonic neuron\cite{nahmias_jstqe_2013} and  a distributed feedback (DFB) laser photonic neuron\cite{nahmias_ipc_2013}. Fig.\,\ref{fig:vcsel_dfb_cross_sections} illustrates the cross sections of the VCSEL and DFB excitable lasers with an embedded SA. Despite their differences, because of the simplicity and generality of our approach, the dynamics of both models can be represented under the same theoretical framework and are two viable candidates for large, integrated photonic neural networks. Each model possesses complementary advantages and disadvantages to the other.

VCSEL photonic neurons occupy small footprints, can be fabricated in large arrays allowing for massive scalability and low power use \cite{koyama_2006}. In addition, the model is amenable to a variety of different interconnection schemes: VCSELs can send signals upward to form 3D interconnects \cite{li_1996}, can emit downward into an interconnection layer via grating couplers \cite{taillaert_2002}, or connect monolithically through intra-cavity holographic gratings\cite{louderback_2004}. An excitable VCSEL with an intra-cavity SA that operates using the same rate equation model described above has already been experimentally realized \cite{barbay_2011}.

DFB laser photonic neurons, in contrast, emit light in the planar direction. Although they have lower wall plug efficiencies, use more power and occupy larger spatial footprints than their VCSEL counterparts, their natural affinity for waveguide coupling and lithographically defined operating wavelengths post growth makes them a strong candidate for integrating photonic neural networks on a single chip. DFB lasers can be coupled into passive waveguides in III-V materials \cite{williams_1990}, can butt couple into waveguides on other materials such as silicon \cite{zhang_taper_2008}, or can be defined within a silicon on insulator (SOI) substrate using techniques such as wafer bonding \cite{Fang:08}.

\begin{table*}[!t]
\renewcommand{\arraystretch}{1.2}
\caption{Typical VCSEL-SA and DFB-SA Excitable Laser Parameters \cite{coldren2011diode,shastri_jqe_2011,nugent_1995,giudice:899,srinivasan_bowers_2011,park_bowers_laser_2007,fang_bowers_2006,nahmias_jstqe_2013}}
\label{tab:laser_neuron_parameters}
\centering
\begin{tabular}{lllll}
	\hline\hline
	{\bfseries Parameter} & {\bfseries Description} & {\bfseries Units} & {\bfseries VCSEL-SA} & {\bfseries DFB-SA}\\
	\hline
	$\eta_{i,a}$ & Active region current-injection efficiency & $-$ & 0.86 & 0.70\\
	$\eta_{i,s}$ & SA region current-injection efficiency & $-$ & 0.86 & 0.70\\	
	$\lambda$ & Lasing wavelength & nm & 850 & 1575\\
	$V_a$ & Active region cavity volume & m$^{\textnormal{3}}$ & 2.4$\times$10$^{-\textnormal{18}}$ & 2.55$\times$10$^{-\textnormal{18}}$\\
	$V_s$ & SA region cavity volume & m$^{\textnormal{3}}$ & 2.4$\times$10$^{-\textnormal{18}}$ & 0.85$\times$10$^{-\textnormal{18}}$\\
	$\Gamma_a$ & Active region confinement factor & $-$ & 0.06 & 0.034\\
	$\Gamma_s$ & SA region confinement factor & $-$ & 0.05 & 0.034\\
	$\tau_a$ & Active region carrier lifetime & ns & 1 & 1\\
	$\tau_s$ & SA region carrier lifetime & ps & 100 & 100\\
	$\tau_\mathrm{ph}$ & Photon lifetime & ps & 4.8 & 2.4\\
	$g_a$ & Active region differential gain/loss & m$^{\textnormal{3}}$s$^{-\textnormal{1}}$ & 2.9$\times$10$^{-\textnormal{12}}$ & 0.97$\times$10$^{-\textnormal{12}}$\\
	$g_s$ & SA region differential gain/loss & m$^{\textnormal{3}}$s$^{-\textnormal{1}}$ & 14.5$\times$10$^{-\textnormal{12}}$ & 14.5$\times$10$^{-\textnormal{12}}$\\
	$n_{0,a}$ & Active region transparency carrier density & m$^{-\textnormal{3}}$ & 1.1$\times$10$^{\textnormal{24}}$ & 1.1$\times$10$^{\textnormal{24}}$\\
	$n_{0,s}$ & SA region transparency carrier density & m$^{-\textnormal{3}}$ & 0.89$\times$10$^{\textnormal{24}}$ & 1.1$\times$10$^{\textnormal{24}}$\\
	$B_r$ & Bimolecular recombination term & m$^{\textnormal{3}}$s$^{-\textnormal{1}}$ & 10$\times$10$^{-\textnormal{16}}$ & 10$\times$10$^{-\textnormal{16}}$\\
	$\beta$ & Spontaneous emission coupling factor & $-$ & 1$\times$10$^{-\textnormal{4}}$ & 1$\times$10$^{-\textnormal{4}}$\\
	$\eta_c$ & Output power coupling coefficient & $-$ & 0.4 & 0.39\\
	$n_{\mathrm{eq},a}$ & Active region equilibrium carrier density & m$^{-\textnormal{3}}$ & 7.86$\times$10$^{\textnormal{15}}$ & 7.86$\times$10$^{\textnormal{15}}$\\
	$n_{\mathrm{eq},s}$ & SA region equilibrium carrier density & m$^{-\textnormal{3}}$ & 7.86$\times$10$^{\textnormal{15}}$ & 7.86$\times$10$^{\textnormal{15}}$\\
	\hline\hline
\end{tabular}
\end{table*}

\section{Results and discussion}\label{sect:results}
\begin{figure*}[t]
	\centering
	\includegraphics[width=3.00in]{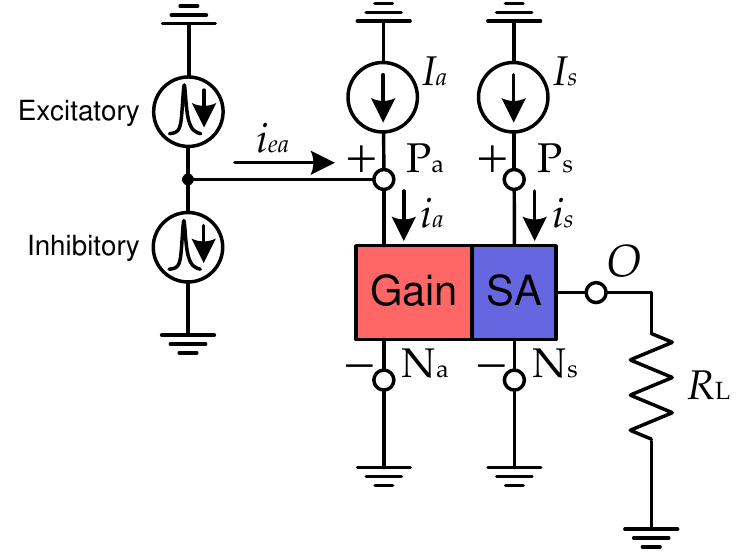}
	\caption{Circuit setup to simulate the laser neuron equivalent circuit model.}
	\label{fig:circuit_setup}
\end{figure*}

We simulate our laser neuron circuit model using the HSPICE circuit simulator from Synopsys\footnote{http://www.synopsys.com/tools/Verification/AMSVerification/CircuitSimulation/HSPICE/}. For our analysis, we consider typical VCSEL-SA and DFB-SA excitable lasers with material and geometrical parameters as given in Table\,\ref{tab:laser_neuron_parameters}. Fig.\,\ref{fig:circuit_setup} depicts the simulation setup used to test the laser neuron equivalent circuit model in Fig.\,\ref{fig:circuit_model}. In the simulation setup, the dc current sources $I_a$ and $I_s$ provide the bias conditions for gain and SA regions, respectively. The pulsed current sources summed as $i_{ea}$, model the excitatory and inhibitory pulses (from other laser neurons), and modulate the gain region. The dummy load $R_L$, enables the measurement of output power from the laser neuron circuit model.

\begin{figure*}[!h]
	\centering
	\subfigure[]{\includegraphics[width=2.4in]{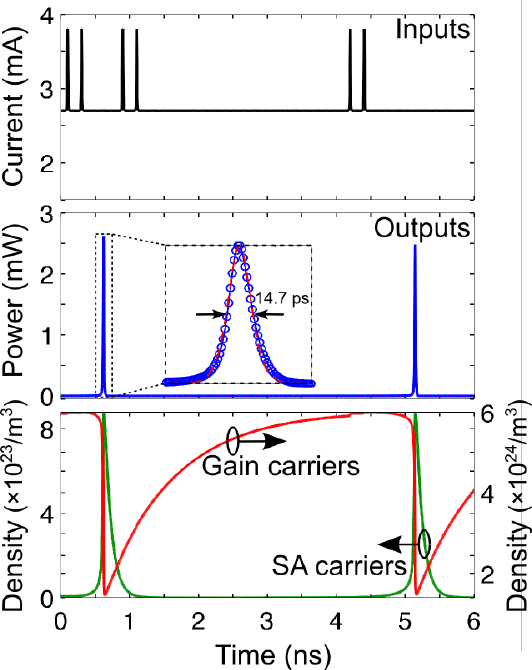}%
		\label{fig:excitatory}}
	\quad
	\subfigure[]{\includegraphics[width=2.4in]{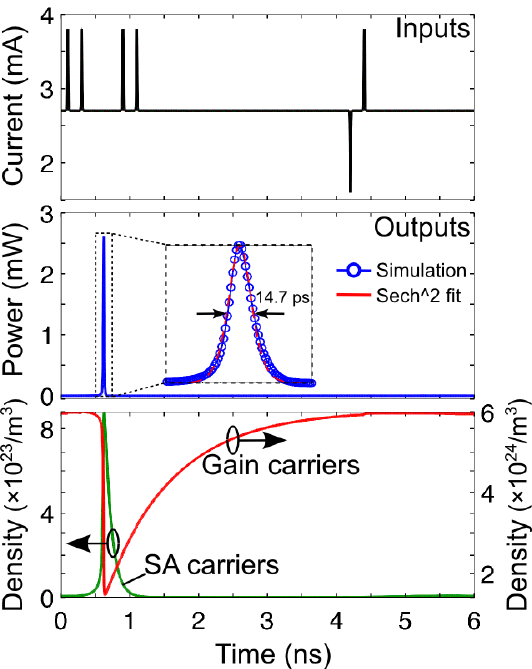}%
		\label{fig:inhibitory}}
	\caption{Simulation of the excitable VCSEL neuron exhibiting neural spiking behavior by selectively modulating the gain through both excitatory (gain enhancement) and inhibitory (gain depletion) pulses. First row: input perturbations to the gain. Second row: output and sech$^2$ fitting curve. Third row: gain region and SA region carrier concentrations.}
	\label{fig:excite_inhibit}
\end{figure*}

As stated earlier, in certain parameter regimes, a simple model of a single-mode laser with an SA section has been proven to be analogous to the equations governing an LIF neuron \cite{nahmias_jstqe_2013,nahmias_ipc_2013}. For such a laser, the active and passive cavities must stay as gain and absorber regions. This means that $I_a>q V_a n_a/\tau_a$ and $I_s<q V_s n_s/\tau_s$, respectively. Fig.\,\ref{fig:excite_inhibit} depicts the simulated excitability of the VCSEL-SA system. This behavior resembles neural spiking behavior. The VCSEL-SA model is biased just below the threshold: gain current, $I_a=2.7$~mA, and SA current, $I_s=0$~mA. In Fig.\,\ref{fig:excitatory} pairs of excitatory spikes are incident on the system at various times. Each excitatory pulse increases the carrier concentration within the gain region by an amount proportional to its energy---gain enhancement. The first pair of pulses causes the laser to fire and emit a pulse, after which a refractory period occurs. During this period, a second pair of pulses is unable to cause the laser to fire. After several nanoseconds, the laser has recovered to its equilibrium state and a third pair of excitatory spikes causes it to fire again. In Fig.\,\ref{fig:inhibitory} an excitatory and inhibitory pair replaces the third pair of excitatory spikes. Inhibition, which decreases the carrier concentration within the gain region---gain depletion---cancels the excitatory activity and prevents the laser from firing.

\begin{figure*}[t]
	\centering
	\subfigure[]{\includegraphics[width=2.3in]{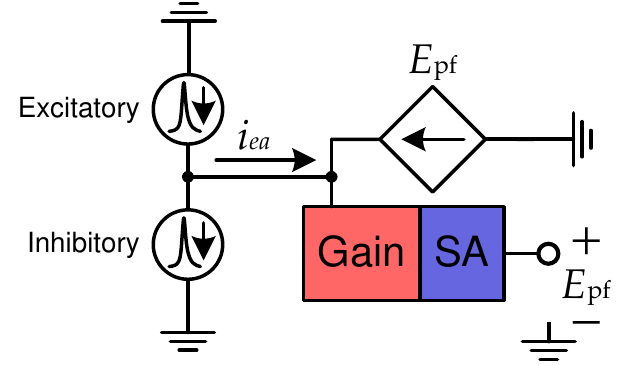}%
		\label{fig:bistability_circuit}}
	\subfigure[]{\includegraphics[width=2.7in]{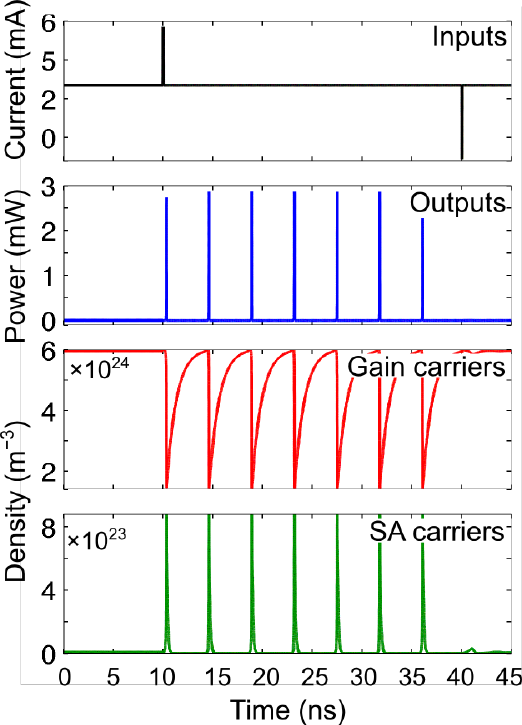}%
		\label{fig:bistability_plot}}
	\caption{(a) Laser neuron circuit setup to investigating bistable dynamics. Note that the dc biasing conditions have been left out for the sake of brevity. (b) Simulation of the excitable VCSEL neuron system exhibiting bistability with connection delays of 4~ns. Top row: input perturbations to the gain. Second row: output power. Third row: gain region carrier concentration. Fourth row: SA region carrier concentration.}
	\label{fig:bistability}
\end{figure*}

A crucial property of dynamical systems that arises out of the formation of hysteric attractors is bistability, which plays an important role in the formation of memory in processing systems \cite{nahmias_jstqe_2013}. Here, the system is recursive rather than feedforward, possessing a network path that contains a closed loop allowing the system to exhibit hysteresis, and it is essentially an extension of an autapse. Fig.\,\ref{fig:bistability_circuit} illustrates an excitable laser whose output is fed back to the input with a delay element. This circuit represents a test of the network's ability to handle recursive feedback. The VCSEL-SA excitable laser model is employed for this simulation with the biasing conditions for the gain and SA regions: $I_a=2.7$~mA (just below threshold) and $I_s=0$~mA. Fig.\,\ref{fig:bistability_plot} shows the result for the simulation. An excitatory pulse input to the laser neuron at $t=10$~ns, initiates the system to settle to a new attractor. The first output pulse is fed back to the input after being delayed by $t_\mathrm{delay}=4$~ns, which initiates another excitatory pulse at the output. This recursive process results in a train of output pulses at fixed intervals before being deactivated by a precisely timed inhibitory pulse at $t=40$~ns. This system successfully maintains the stability of self-pulsations. The system is also capable of stabilizing to other states, including those with multiple pulses or different pulse intervals and thus acts as an optical pattern buffer over longer time scales \cite{nahmias_jstqe_2013}.

Note, since the output of the laser neuron is light represented in the circuit model as a voltage and its input is a current, a voltage-controlled current source is employed for the feedback connection. A practical implementation for interconnecting laser neurons requires a photodetector at its input which can couple the optical outputs of other laser neurons and electrically modulate the gain section of the laser neuron. We recently proposed such an excitable laser and photodetector system that can emulate both a LIF neuron and a synaptic variable, completing a computational paradigm for scalable optical computing \cite{nahmias_ipc_2013}. Here, two photodetectors receive optical pulses (excitatory and inhibitory) from a network and are subtracted passively by a push-pull wire junction. The resulting photocurrent signal conducts over a short wire to modulate the laser gain section. Dynamics introduced by the photodetectors are analogous to synaptic dynamics governing the concentration of neurotransmitters in between signaling biological neurons. Future work will include a first-order low-pass filter to model the photocurrent flow in photodiodes and neural synaptic dynamics. The conversion between optical and electronic domains also restricts the propagation of optical phase noise and the need for direct wavelength conversion, thus eliminating two major barriers facing scalable optical computing \cite{nahmias_ipc_2013}. 

\begin{figure*}[t]
	\centering
	\subfigure[]{\includegraphics[width=2.5in]{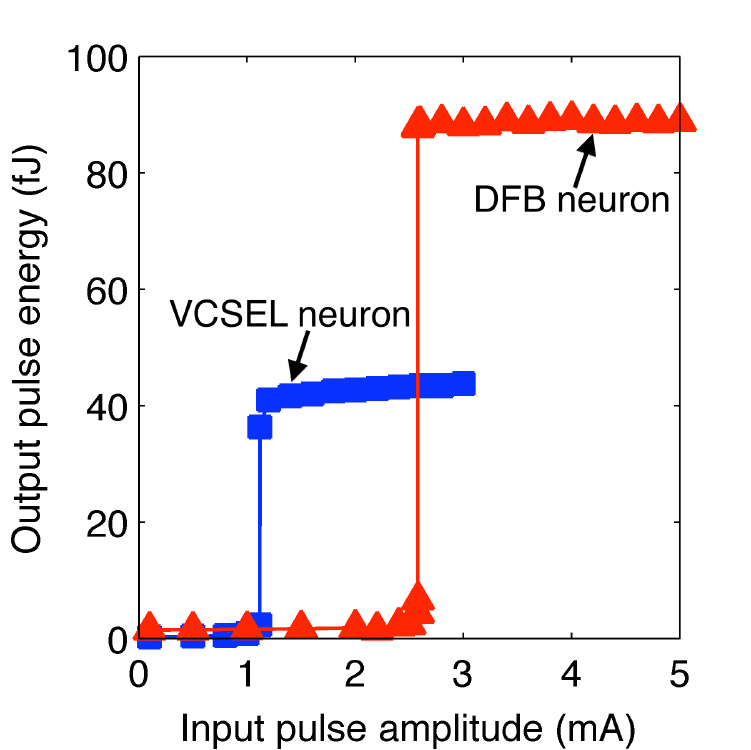}%
		\label{fig:threshold_energy}}
	\subfigure[]{\includegraphics[width=2.5in]{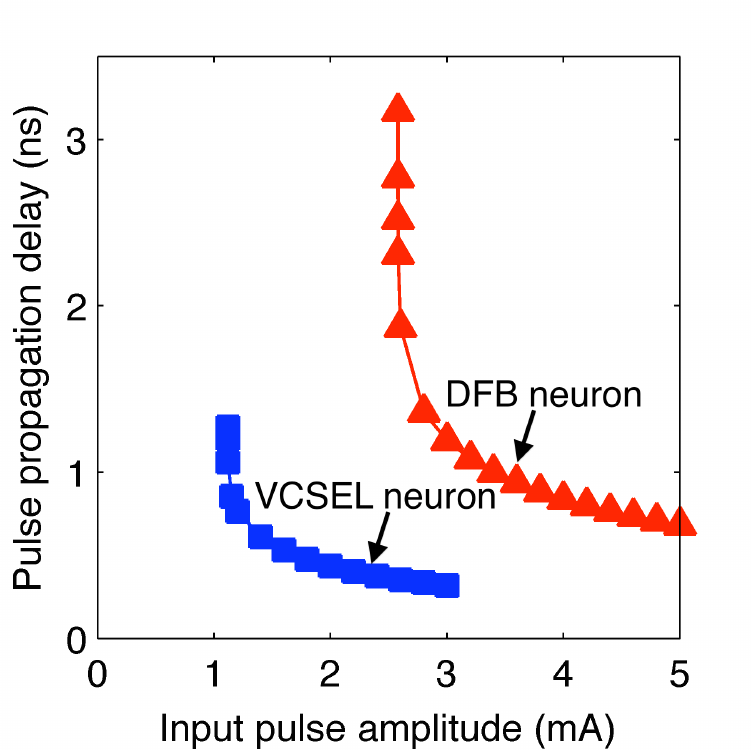}%
		\label{fig:threshold_delay}}
	\caption{Characteristics of the excitable laser neurons biased just below threshold (LIF VCSEL-SA: $I_a=2.7$~mA and $I_s=0$~mA; LIF DFB-SA: $I_a=16.45$~mA and $I_s=0$~mA) to a single impulse . (a) Energy transfer function. (b) Decision latency as a function of the input pulse energy.}
	\label{fig:threshold_plots}
\end{figure*}

We now consider the response of the excitable laser neurons to a single impulse over a range of amplitudes. The underlying feedback dynamics of the laser cavity yield an ideal all-or-nothing energy transfer function that resembles a step as shown in Fig.\,\ref{fig:threshold_energy}. More than a logic level buffer that makes input assumptions, the output of the laser neuron is binary for any analog input. For neural systems, a binary all-or-nothing pulse output is critical to ensure amplitude robustness to channel noise. Other systems, including analog-to-digital converters and comparators, also rely on an all-or-nothing response. When used as a thresholder, the laser neuron has several uncommon advantages over prior photonic thresholding technologies that do not use laser cavity feedback \cite{tait_dream_2013,kravstov_2007}. Favorable properties include an all-or-nothing response, clean pulse generation regardless of input pulse shape, and low threshold amplitude less than a few~mAs.

In phase space, the decision of output pulse or no pulse is determined by an attractive manifold (a.k.a separatrix), which separates the basin of attraction of the at rest steady-state from a large transient trajectory. There are no intermediate output possibilities, but the decision latency is greater when the input causes the dynamical state to fall close to the separatrix as depicted in Fig.\,\ref{fig:threshold_delay}. This near-threshold spike latency, characteristic of separatrix dynamics, is also observed in the Hodgkin--Huxley model \cite{hodgkin_huxley}. While variable latency can lead to jitter in synchronous systems, it has also been proposed as a mechanism for converting continuous amplitude to spike-timing codes in certain neural systems such as the retina \cite{izhikevich_2006}. Further study will focus on the effects of noise on the energy transfer function and decision latency.

\section{Conclusion}\label{sect:conclusion}
There has been a recent surge in unconventional computing paradigms that leverage the underlying physics of devices to breach the limitations inherent in traditional von Neumann architectures and CMOS device technologies. Inspired by biological neural networks, cognitive computers could outperform current technology in both complexity and power efficiency. The computational primitive for such a platform is the well-studied and paradigmatic spiking neuron. A photonic realization of spiking neuron dynamics---laser neurons---harnesses the high-switching speeds, wide communication bandwidth of optics, and low cross-talk achievable in photonics, making it well suited for an ultrafast spike-based information scheme. 

In this paper, we have proposed an equivalent circuit model for laser neurons based on excitable lasers with an SA and direct gain injection. We show that by mapping the laser neuron rate equations into a circuit model, SPICE analysis can be used as an efficient and accurate engine for numerical calculations. Due to the strongly-coupled photon and carrier interactions, the laser neuron rate equations can be only solved using numerical methods such as the Runge--Kutta method. In the same way, the non-perturbative dynamics of neuron biophysics prevent any analytical solution to their behavior. Hodgkin and Huxley were the first to take an approach of mapping the underlying physics present in the neuron to equivalent circuit representations, in order to describe and simulate them within a universal and powerful engineering framework. We take a parallel approach in mapping the key physics of excitable lasers to an equivalent circuit, in hopes of establishing a foundation of comparable utility for laser neuron research.

By leveraging the modularity advantages of a circuit abstraction, SIMPEL will serve as a powerful tool in future studies. We have applied the reported model to different excitable laser types (i.e. VCSEL-SA and DFB-SA) and employed it to study signal-processing behaviors, including excitability with excitatory and inhibitory pulses, binary all-or-nothing response, and bistability. As a next step, we will investigate the implementation of spike-timing-dependent plasticity (STDP)---one of the most important algorithms for spike-based learning \cite{abbott00,savin_2010}---with our model. Optical STDP operating on unprecedented time scales is potentially useful for applications including coincidence detection, sequence learning, path learning, and directional selectivity in visual response. Further work could also investigate power consumption, temperature, and noise, for studying the architecture of an interconnected network of laser neurons at the system level that could have a wide range of applications in high-performance computing, adaptive control, and RF spectrum processing.

\section*{Acknowledgments}
The work of B.J.S was supported by the Banting Postdoctoral Fellowship administered by the Government of Canada through the Natural Sciences and Engineering Research Council of Canada. The work of A.N.T and M.A.N was supported by the National Science Foundation Graduate Research Fellowship Program.

\end{document}